\documentclass[ reprint, superscriptaddress, aps,prl,lengthcheck, twocolumn, showpacs]{revtex4}%
\usepackage{graphicx}
\usepackage{dcolumn}
\usepackage{bm}
\usepackage{epstopdf}
\usepackage{amsmath}
\usepackage{amsfonts}
\usepackage{amssymb}
\providecommand{\U}[1]{\protect\rule{.1in}{.1in}}
\begin{document}
\title{The Photonic Band theory and the negative refraction experiment of
metallic helix metamaterials}
\author{Chao Wu}
\affiliation{Physics Department, Tongji University, Shanghai 200092,
China}
\author{Hongqiang Li}
\email{hqlee@tongji.edu.cn} \affiliation{Physics Department, Tongji
University, Shanghai 200092, China}
\author{Zeyong Wei}
\affiliation{Physics Department, Tongji University, Shanghai 200092,
China}
\author{Xiaotong Yu}
\affiliation{Physics Department, Tongji University, Shanghai 200092,
China}
\author{C.T. Chan}
\affiliation{Department of Physics, Hong Kong University of Science
and Technology, Clear Water Bay, Kowloon, Hong Kong, China}

\begin{abstract}
We develop a theory to compute and interpret the photonic band
structure of a periodic array of metallic helices for the first
time. Interesting features of band structure include the ingenuous
longitudinal and circularly polarized eigenmodes, the wide
polarization gap [Science \textbf{325}, 1513 (2009)], and the
helical symmetry guarantees the existence of negative group velocity
bands at both sides of the polarization gap and band crossings
pinned at the zone boundary with fixed frequencies. A direct proof
of negative refraction via a chiral route [Science \textbf{306},
1353 (2004)] is achieved for the first time by measuring
Goos-hanchen shift through a slab of three dimensional bona fide
helix metamaterial.

\end{abstract}


\maketitle

With a pitch angle accounting for mirror asymmetry, a helix is a
prototype element of chiral media. As a pioneer study, the
electromagnetic(EM) activity from randomly dispersed metallic
helices was observed in 1914 \cite{1}. After the prediction on
negative refraction via chiral route \cite{2}, chiral metamaterials
have been arousing more and more attentions recently on negative
refraction index \cite{3,4,5,6}, strong optical activity
\cite{7,8,9} and circular dichroism \cite{10,11,12,13} as well. In
the most of previous studies, chiral metamaterials are typically
realized as one or two layers of discrete chiral resonators and the
negative refractive index is deduced indirectly from retrieval
procedures \cite{3,4,5,6} assuming the applicability of effective
medium theory \cite{14,15,16,17,18}. Recently, the Gold helix
metamaterial with a few pitches along the helical axis are
successfully fabricated in THz regime with a wide polarization gap
measured experimentally \cite{19}, which also raised a question
whether the effective medium theory is enough for the description of
chiral metamaterials. To the best of our knowledge, most previous
studies adopt effective medium description and well-defined EM
parameters for theoretical analysis, and empirically interpret the
exotic properties in a numerical stage. While there is no any
``first-principle'' investigation starting from the real structure
of chiral metamaterial. Also it is worth noting that in all previous
studies, negative refraction indices, retrieved from the
transmission and reflection spectra under normal incidence, do not
give any direct proof for negative refraction. The negative
refraction via chiral route hasn't really achieved before.

In this paper, we theoretically and experimentally investigate the
properties of a three-dimensional metamaterial made with a square
array of metallic helices. We develop a photonic band theory for
helix metamaterial by combining Multiple Scattering Theory (MST)
\cite{20,21} with Sensiper's solution for a single helix \cite{22}.
The theory enables us to identify the longitudinal and circularly
polarized eigenmodes, the formation of the wide polarization gap,
negative group velocity bands at both sides of the gap and related
fine features. We also demonstrate a proof of principle experiment
by measuring negative Goos-hanchen shift to directly verify the
negative refraction from helix metamaterial for the first time.
\begin{figure}[pb]
\centerline{\includegraphics[width=7.5cm]{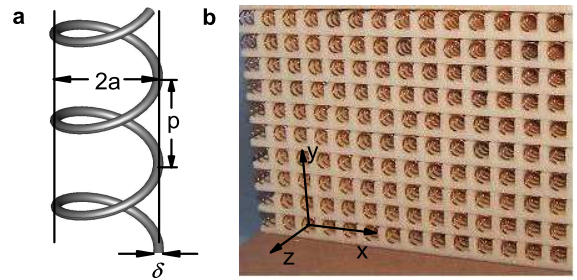}} \caption{(a)
The schematic picture of a metallic helix unit and (b) photo of a
square array of metallic helices.} \label{fig1}
\end{figure}
\begin{figure*}[pth]
\centerline{\includegraphics[width=16cm]{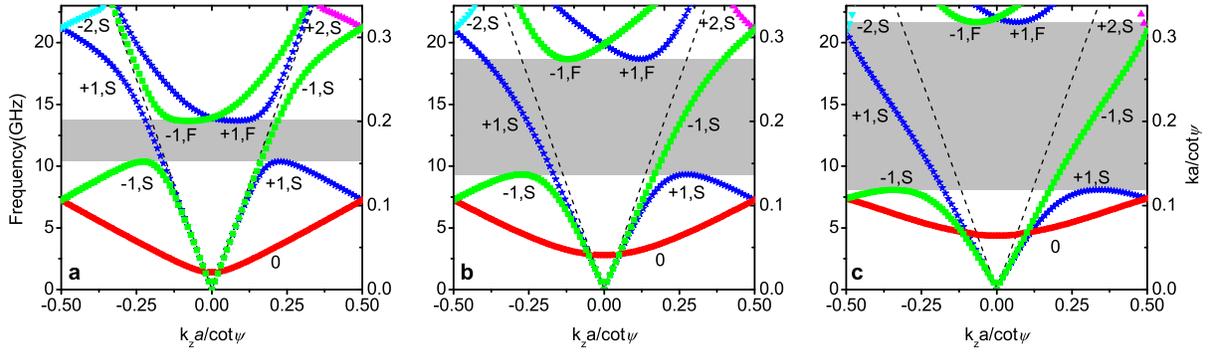}} \caption{The
photonic band structures of the helix crystal along the helix axis
for three lattice constants (a) $d=20$ mm, (b) $d=11$ mm, (c) $d=8$
mm. The pitch $p=4.4$ mm, the radius $a=3.3$ mm, and diameter of
metallic wires $\delta=0.8$ mm.}%
\label{fig2}%
\end{figure*}

Figure \ref{fig1} shows a photograph of a sample of the
three-dimensional metamaterial comprising a square array of
right-handed (RH) helices. Oriented along the $z$ axis, the metallic
helices form a square array in the \textit{xy} plane with a lattice
constant of $d=11$ mm. Figure \ref{fig1}(a) shows the schematic
picture of a single helix unit, which has a pitch of $p=4.4$ mm,
radius $a=3.3$ mm, and diameter of metallic wires $\delta=0.8$ mm.
We define a pitch angle $\psi$ by $\cot \psi =2\pi a/p$. We note
that a helix comes back to itself after being translated by a
distance of $\Delta z$ and being rotated simultaneously by a angle
of $2\pi \Delta z/p$ (for RH helix) or $-2\pi \Delta z/p$ [for
left-handed (LH) helix], and thus physical entities associated with
the helix should satisfy the helical symmetry condition
\begin{equation}
\label{eq1}
U\left( {\rho ,\phi ,z} \right)=U\left( {\rho ,\phi \pm \frac{2\pi \Delta
z}{p},z+\Delta z} \right)
\end{equation}
with +/- sign for the RH/LH helix respectively \cite{22,23}. As the
helical system is also periodic along the helical axis, the field
components for an RH helix system can be expanded by functions of
the form
\begin{equation}
\label{eq2} \psi _n \left( {\rho ,\phi ,z} \right)=e^{ik_z z}F_n
\left( \rho \right)e^{-in\phi }e^{i\frac{2n\pi }{p}z}
\end{equation}
where $k_z$ is the Bloch wavevector along $z$ axis. The angular term
should be $e^{in\phi }$ instead if the helix is LH. The radial
function $F_n \left( \rho \right)$ obeys the Helmholtz differential
equation, and can be expressed in terms of modified Bessel functions
$I_{n}$ and $K_{n}$. We follow the Sensiper approach \cite{22} to
impose the assumption about uniformly distributed surface current
flow along the metal wires. Under that assumption, the boundary
continuity conditions require that the local electric field on metal
wires must be perpendicular to the line of metal wire. Thus we
derive an eigenvalue equation of helix array by MST as
\begin{eqnarray}
\label{eq3} \sum\limits_n \left[ \left( {k_z^2
a^2-k^2a^2+\frac{n^2k^2}{\tau_n^2 }\cot ^2\psi } \right)y_n I_n(\tau _n a) \right. \nonumber \\
+k^2a^2\cot ^2\psi z_n I_n '(\tau _n a) \bigg]\frac{R_n }{x_n \tau
_n } =0
\end{eqnarray}
where $\tau _n=\left[(k_z+\frac{2\pi
n}{p})^2-k^2\right]^{\frac{1}{2}},$ $R_n =\frac{\sin(n\pi \delta
/p)}{n\pi \delta /p},$ $y_n = K_n ( \tau _n a)+( -1)^n\sum_l S_{l-n}
( \tau _n )I_l ( \tau _n a), \qquad \qquad$ $ z_n =K_n '(\tau _n
a)+(-1)^n\sum_l S_{l-n}(\tau _n )I_l '( \tau _n a),\qquad \qquad$
$x_n =y_n I_n '(\tau _n a)-z_n I_n (\tau _n a),\qquad \qquad \qquad
\qquad \qquad$ $\mbox{and} \quad S_l ( \tau )=\sum\limits_{q\ne 0}
K_l (\tau R_q )e^{il\phi _q }e^{i \textbf{\textit{k}}_\textbf{i}
\cdot \textbf{R}_\textbf{q} }$ is a lattice sum running over the
nodes $\left( {R_q ,\phi _q } \right)$ of the square lattice in
cylindrical coordinates with $\textbf{\textit{k}}_\textbf{i}$ being
the transverse component of the wavevector \textbf{\textit{k}} in
the air, and $I_n '(x)$, $K_n '(x)$ satisfy to $I_n '(x)=dI_n
(x)/dx$, $K_n '(x)=dK_n (x)/dx$.

\begin{figure}[ptbh]
\centerline{\includegraphics[width=8.5cm]{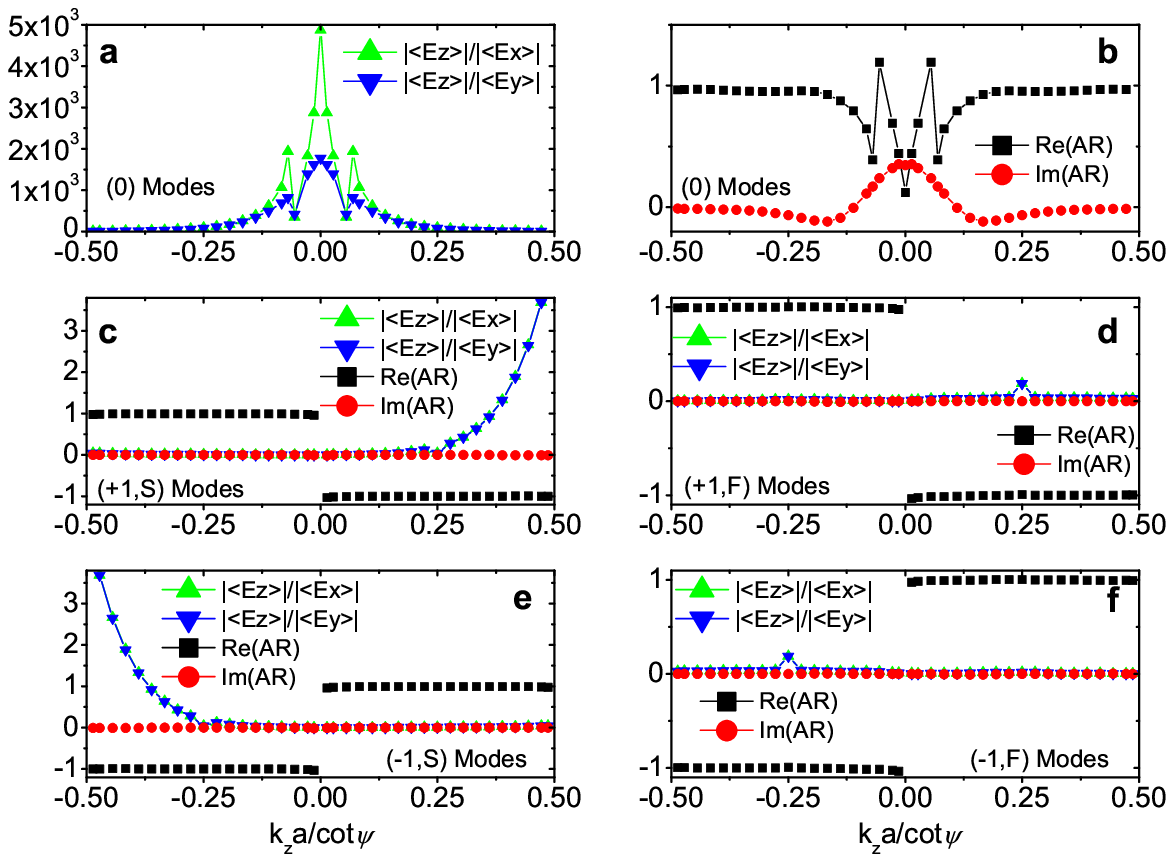}}\caption{Eigenmode
analysis by calculating the ratios $\mbox{AR}=\left\langle {E_x }
\right\rangle k_z /\left\langle {iE_y } \right\rangle \left| {k_z }
\right|$, $\left| {\left\langle {E_z } \right\rangle }
\right|/\left| {\left\langle {E_x } \right\rangle } \right|$ and
$\left| {\left\langle {E_z } \right\rangle } \right|/\left|
{\left\langle {E_y } \right\rangle } \right|$. (a) and (b) for $n=0$
modes; (c) (+1,S) modes; (d) (+1,F) modes; (e)
(-1,S) modes; (f) (-1,F) modes.}%
\label{fig3}%
\end{figure}

The photonic band structures, computed with the combined technique
above stated, afford us an intuitive understanding about optics of
helix metamaterials. In the helical structure, there is always a
$\pi/2 $ or $-\pi/2 $ phase difference between the radial and the
angular components for both the electric field and magnetic field,
implying that the eigenmodes are ingenuous left-handed or right
handed circularly polarized (LCP or RCP). This can be checked by
examining the field solutions written in the Sensiper form\cite{22}
, or by examining the solution from commercial numerical solver CST.
Figures \ref{fig2}(a), \ref{fig2}(b) and \ref{fig2}(c) show the band
structure of the helix metamaterials along the helix axis for three
lattice constants $d =20$ mm, 11 mm and 8 mm respectively. We label
the eigenmodes by its dominant term in Eq. (\ref{eq2}). For example,
the (-1,S) modes (blue stars in Fig. \ref{fig2}) have $n= -1$ term
as the dominant term and ``S'' stands for a ``slow mode'' below the
light line, and we use the subscript ``F'' for a mode inside the
light cone. The lowest frequency branch (red circles), labeled as
(0) in Fig. \ref{fig2}, has a strong longitudinal component and this
mode picks up a LH character as $k_z$ increases. It goes to a
longitudinal mode with a finite frequency at Brillouin zone (BZ)
center ($k_z=0)$. Both the electric field and magnetic field are
essentially parallel to the helical axis. It is evident from Fig.
\ref{fig2} that the inter-helix coupling pushes the longitudinal
mode to higher frequencies. In the limit $d\to \infty $, it goes to
zero frequency. The polarization of an eigenmode is analyzed by the
ratios $\left| {\left\langle {E_z } \right\rangle } \right|/\left|
{\left\langle {E_x } \right\rangle } \right|$, $\left| {\left\langle
{E_z } \right\rangle } \right|/\left| {\left\langle {E_y }
\right\rangle } \right|$ and $\mbox{AR}=\left\langle {E_x }
\right\rangle k_z /\left\langle {iE_y } \right\rangle \left| {k_z }
\right|$, where the spatial average, $\left\langle \cdots
\right\rangle $, is taken inside a unit cell. Figure \ref{fig3}
clearly indicate that the polarization of $n=0, \pm 1$ eigenmodes
are longitudinal and/or LCP, RCP exclusively. Eigenmode analysis and
numerical transmission simulation employing a finite thickness slab
showed that the $n = -1 $ modes with positive group velocity (blue
stars in Fig. \ref{fig2}) couple to incident plane wave with
opposite handedness as the helix, while the $n = +1$ modes with
positive group velocity (green squares in Fig. \ref{fig2}) couple to
incident waves of the same handedness as the helix. In general, a
mode couples to incident wave of the same (opposite) handedness as
the helix if $n\cdot k_z
>0(<0)$.

An important feature of the band structure is the existence of a
wide polarization gap (shaded in grey in Fig. \ref{fig2}) that only
allows incident waves of opposite handedness to pass through. As
such, a RH helix array has a right-handed circularly polarized (RCP)
gap. The gap grows wider for a higher helix filling ratio (smaller
$d$). Such wide polarization gaps have been experimentally
demonstrated for a thin slab of gold helices in IR frequencies
\cite{19,24}. The lower edge of the polarization gap is pinned at
the frequency at which the (+1,S) mode attains zero group velocity.
The (+1,S) mode is the result of the hybridization between the free
photons (riding on the light line) and a mode guided on the helix
which is strongly back-scattered by the periodicity. As the (+1,S)
mode is guided on the helices, the frequency is only moderately
affected by the inter-helix coupling. The upper edge of the
polarization gap is pinned at the frequency at which the (+1,F) mode
has zero group velocity. The (+1,F) mode is not a guided mode on a
single helix. It exists as an eigenmode only in the periodic array
and is sensitive to the volume available between the helices and the
mode is squeezed to higher frequencies at a higher helix filling
ratio, leading to a much bigger polarization gap for smaller values
of the lattice constant ($d$). In order words, the lower edge of the
gap is primarily determined by the helix parameters ($a$ and $p$)
along the helical axis, while the upper edge is primarily determined
by the structural parameter perpendicular to the helical axis
(lattice constant $d$).

Another interesting feature of the band structure is the emergence
of the negative group velocity bands at both sides of the
polarization gap, which is different from the previous notions that
the negative refraction only happens above the resonant gap. Both
the high frequency (+1,F) and the lower frequency (+1,S) branch
exhibit negative group velocities. The (+1,S) branch has negative
group velocity after reaching a maximum frequency that pins the
lower edge of the polarization gap. Concomitant with the negative
refraction bands in slow mode, one can see that there are no band
gap but band crossings at the BZ boundary ($k=\pi /p$) and the
degenerate modes are pinned at fixed frequencies that are nearly
independent of lattice constant $p$. We note that according to Eq.
\ref{eq2} Bloch modes that differ by $\Delta k=2\pi /p$ will be
orthogonal because of orthogonality of the angular phase factor. And
the (0) branch degenerates with the (+1,S) branch at zone center,
giving rise to a negative group velocity band below the polarization
gap. This is essentially a consequence of backward waves guided on
metallic helix satisfying to helical symmetry. As a proof, a
comparative study shows that the negative dispersion band disappears
as well as other unique features, such as the longitudinal mode,
band crossing etc. when the helix metamaterial is cut in a
``discrete'' form by inserting an 0.4mm air gap at each
period(\emph{p}).

We performed negative refraction measurements inside an anechoic
chamber through a slab of the helix metamaterial with the
aforementioned geometric parameters [the band structure is shown in
Fig. \ref{fig2}(b)]. Helix metamaterial samples are fabricated by
periodically embedding the clockwise metallic helices in a
polyurethane foam slab. The polyurethane foam slab is lossless with
$\varepsilon \approx 1$. The sample slab contains $15\times 11$
metallic helices, each having 140 periods along helical axis ($z$
axis).
\begin{figure*}[ptb]
\centerline{\includegraphics[width=15cm]{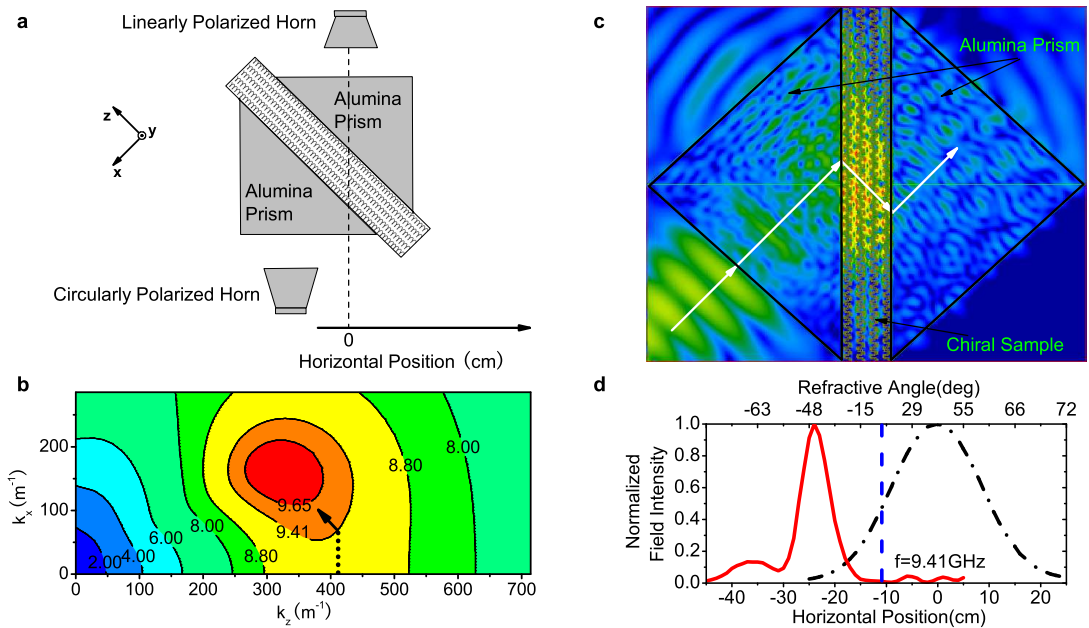}} \caption{(a)
Schematic of the experimental setup for negative refraction. (b)
Computed equi-frequency surface (EFS)for the (+1,S) branch. The
arrow refers to the direction of the refracted waves at 9.41 GHz
under an incident angle at 45\r{ }. (c) FDTD simulations of negative
refraction. In calculation, an RCP Gaussian beam with a frequency f
= 9.41 GHz is normally impinging on the left prism. The chiral
sample has $4\times5$ helices in the \textit{xy} plane and 45
periods along the \textit{z} axis. The direction of energy flow is
indicated by the arrow. (d) Measured electric field intensity as a
function of the horizontal position of the circularly polarized horn
receiver, with (red) or without (black) the chiral sample and
alumina prisms at 9.41 GHz. The two curves are normalized such that
the magnitude of both peaks is unity. The blue dashed line at
horizontal position of $-11$ cm refers to Goos-Hanchen shift with
respect to 0\r{ } refracted angle.} \label{fig4}
\end{figure*}

Computed equi-frequency surface (EFS) for the (+1,S) branch [see
Fig. \ref{fig4}(b)] and (+1,F) branch (not shown) demonstrate that
negative refraction can be achieved at both sides of the
polarization gap. Here we try to realize the negative refraction at
the lower edge of the gap, which is not found or predicted in other
systems before. As the ($\pm $1,S) modes lie below the light line,
we can excite the ($\pm $1,S) modes by prism coupling techniques to
penetrate a Gaussian beam into the sample slab [see Fig.
\ref{fig4}(a)], and estimate the refractive angle quantitatively by
measuring the Goos-Hanchen shift of the beam. Figure \ref{fig4}(a)
illustrates the schematic configuration of our experimental setup.
Two isosceles right-angled triangular alumina prisms ($\varepsilon
_r =8.9$) are placed so that they touch the sample slab at both
sides and a Gaussian beam is normally incident in $xz$ plane to the
air-prism interface from a linearly polarized horn emitter
(operating at 8.2-12.4 GHz with a gain factor of 24.6 dB), ensuring
an incident angle of 45\r{ } from alumina to metamaterial. The local
field intensity is measured by the LCP/RCP horn receiver as a
function of the horizontal position in a precision of 1 mm per step.
The Goos-Hanchen shift of the outgoing beam is quantified by
measuring the peak position at the interface of prism. The
coordinate origin in the horizontal position is aligned to the
position of the horn emitter, marked by the dashed vertical line in
Fig. \ref{fig4}(a). The negative refraction is measured from 9.18
GHz to 9.48 GHz with a refraction angle from $-17.44$\r{ } to
$-50.11$\r{ }, which is in good agreement with the computed
equi-frequency surface (EFS) shown in Fig. \ref{fig4}(b). The solid
line in Fig. \ref{fig4}(d) presents the spatial profile of local
field intensity measured at 9.41 GHz. A peak value is measured at
the horizontal position of $-24$ cm, corresponding to a refraction
angle of $-46.45$\r{ }, roughly equal to $-45.82$\r{ } estimated by
EFS analysis. Thus negative refraction below the polarization gap is
verified experimentally. Negative refraction is also qualitatively
verified by finite-difference-in-time-domain (FDTD) simulations
shown in Fig. \ref{fig4}(c). We note from Fig. \ref{fig4}(b) that in
the frequency range of our interest, the dispersion is not only
negative along $k_z$, but it is also negative along $k_x$ due to
Bragg scattering.

In summary, we witness the mathematical beauty of wave propagation in
metallic helix metamaterial by analytical resolved photonic band structure.
There are negative bands both above and below the polarization gap. Negative
refraction on the low frequency branch is demonstrated directly for the
first time by a Goos-hanchen shift experiment. We note that the optics of
helix metamaterial is governed by the collective selection on the guided
modes on helices satisfying to helical symmetry and (evanescent) Bragg
scatterings within helix lattice and in passing there is no easy way to
describe the phenomena using effective medium parameters. This work was
supported by NSFC (No. 10974144, 60674778), HK RGC grant 600308, the
National 863 Program of China (No.2006AA03Z407), NCET (07-0621), STCSM and
SHEDF (No. 06SG24). We thank for M. Wegener and J. Gansel for helpful
discussions.


\begin{thebibliography}{99}                                                                                               %

\bibitem {1}I. V. Lindell, A. H. Sihvola, and J. Kurkijarvi, IEEE Trans Ant. {\&} Prop.  \textbf{34}, 24 (1992).

\bibitem {2}J. B. Pendry, Science \textbf{306}, 1353 (2004).

\bibitem {3}E. Plum et al., Phys. Rev. B \textbf{79}, 035407 (2009).

\bibitem {4}S. Zhang et al., Phys. Rev. Lett. \textbf{102}, 023901 (2009).

\bibitem {5}J. Zhou et al., Phys. Rev. B \textbf{79}, 121104(R) (2009).

\bibitem {6}M. C. K. Wiltshire, J. B. Pendry, and J. V. Hajnal, J. Phys.: Condens. Matter \textbf{21}, 292201 (2009).

\bibitem {7}E. Plum et al., Appl. Phys. Lett. \textbf{90}, 223113(2007).

\bibitem {8}H. Liu et al., Phys. Rev. B \textbf{76}, 073101 (2007).

\bibitem {9}T. Q. Li et al., Appl. Phys. Lett. \textbf{92}, 131111 (2008).

\bibitem {10}S. L. Prosvirnin, and N. I. Zheludev, Phys. Rev. E \textbf{71}, 037603 (2005).

\bibitem {11}A. V. Krasavin et al., Appl. Phys. Lett. \textbf{86}, 201105 (2005).

\bibitem {12}V. A. Fedotov et al., Phys. Rev. Lett. \textbf{97}, 167401 (2006).

\bibitem {13}E. Plum et al., Phys. Rev. Lett. \textbf{102}, 113902 (2009).

\bibitem {14}N. Engheta, D. L. Jaggard, and M. W. Kowarz, IEEE Trans Ant. {\&} Prop. \textbf{40}, 367 (1992).

\bibitem {15}S. Tretyakov et al., J. Electron Waves and Appl. \textbf{17}, 695 (2003).

\bibitem {16}T. G. Mackay, and A. Lakhtakia, Phys. Rev. E \textbf{69}, 026602 (2004).

\bibitem {17}C. Monzon, and D. W. Forester, Phys. Rev. Lett. \textbf{95}, 123904 (2005).

\bibitem {18}Q. Cheng, and T. J. Cui, Phys. Rev. B \textbf{73}, 113104 (2006).

\bibitem {19}J. K. Gansel et al., Science \textbf{325}, 1513 (2009).

\bibitem {20}S. K. Chin, N. A. Nicorovici, and R. C. McPhedran, Phys. Rev. E \textbf{49}, 4590 (1994).

\bibitem {21}N. A. Nicorovici, R. C. McPhedran, and L. C. Botten, Phys. Rev. E \textbf{52}, 1135 (1995).

\bibitem {22}S. Sensiper, Proc. IRE \textbf{43}, 149 (1955).

\bibitem {23}J. R. Pierce, Proc. IRE \textbf{35}, 111 (1947).

\bibitem {24}J. K. Gansel et al., Opt. Exp. \textbf{18}, 1059 (2010).

\end{thebibliography}
\end{document}